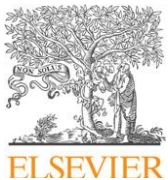
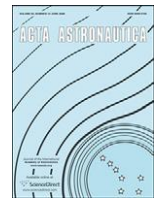

# SETI Reloaded: Next Generation Radio Telescopes, Transients and Cognitive Computing

M.A. Garrett [a,b]

[a] ASTRON, the Netherlands Institute for Radio Astronomy, Postbus 2, 7990 AA, Dwingeloo, The Netherlands
[b] Leiden Observatory, Leiden University, PO Box 9513, 2300 RA Leiden, The Netherlands

### abstract

The Search for Extra-terrestrial Intelligence (SETI) using radio telescopes is an area of research that is now more than 50 years old. Thus far, both targeted and wide-area surveys have yet to detect artificial signals from intelligent civilisations. In this paper, I argue that the incidence of co-existing intelligent and communicating civilisations is probably small in the Milky Way. While this makes successful SETI searches a very difficult pursuit indeed, the huge impact of even a single detection requires us to continue the search. A substantial increase in the overall performance of radio telescopes (and in particular future wide-field instruments such as the Square Kilometre Array - SKA), provide renewed optimism in the field. Evidence for this is already to be seen in the success of SETI researchers in acquiring observations on some of the world's most sensitive radio telescope facilities via open, peer-reviewed processes. The increasing interest in the dynamic radio sky, and our ability to detect new and rapid transient phenomena such as Fast Radio Bursts (FRB) is also greatly encouraging. While the nature of FRBs is not yet fully understood, I argue they are unlikely to be the signature of distant extra-terrestrial civilisations. As astronomers face a data avalanche on all sides, advances made in related areas such as advanced Big Data analytics, and cognitive computing are crucial to enable serendipitous discoveries to be made. In any case, as the era of the SKA fast approaches, the prospects of a SETI detection have never have been better.

## 1. Introduction

The Search for Extra-terrestrial Intelligence (SETI) is a field of research that is now more than 50 years old [1, 2]. Thus far, no compelling detection of an artificial radio signal from another civilization has been made despite various ad hoc, targeted and systematic surveys being performed e.g. [3]. In this paper, I argue that at any given time, the chances that there are two civilisations co-existing in the Milky Way, and able to communicate with each other is probably relatively small - the vast distances between planetary systems, together with the huge range of relevant timescales (e.g. the durability of planet production in the Milky Way, compared to the lifetime of a radio communicating civilisation), all tend to work against each other. Nevertheless, while the probability of SETI success may be small, it is not zero, and the recent emergence of new SETI searches using the latest radio astronomy facilities are well justified. Indeed, considering the potential wide-spread impact of a SETI detection on all mankind, and noting that our capabilities to perform advanced SETI surveys currently improve exponentially with time, the imperative to continue and indeed renew and refine our efforts in this field is vital.

## 2. The SETI challenge

A simplified form of the Drake equation [4] provides a good starting point in order to estimate the magnitude of the challenge SETI researchers face. In its simplified form, the Drake equation is separated into its astronomical, biological and sociological terms. The current number of co-existing radio-communicating civilisations is then estimated as:

$$N \sim R_{ASTRO}\, f_{BIO}\, L$$

where $R_{ASTRO}$ represents the number of habitable planets currently being produced in the Milky Way each year, $f_{BIO}$ represents the product of all chemical, biological and sociological probability factors leading to the development of a technological civilisation, and $L$ is the average duration of the radio-communicating phase of a civilisation.

Today, $R_{ASTRO}$ is known rather well, although the exact numbers can of course be argued over. For the sake of simplicity, we estimate that the current value of $R_{ASTRO}$ is ~ 1 planet per year. Similarly, if we consider the most highly *optimistic* estimate for the biological factors ($f_{BIO} \sim 1$), then the simplified Drake equation becomes:

$$N \sim L$$

In other words, if the average duration of the radio-communicating phase of a civilisation is (for example) 3000 years, we expect about 3000 civilisations to currently co-exist with each other in the Milky Way. Assuming (again for simplicity), that these civilisations are randomly spread throughout the Galaxy, the typical distance between co-existing, communicating extra-terrestrial civilisations ($D_{CCETI}$) is given by [see 4]:

$$D_{CCETI} \sim 2 R_{Galaxy}/\sqrt{N}$$

For $N \sim 3000$, and $R_{Galaxy} \sim 10$ kpc (~ 30000 light years), the typical distance between radio communicating civilisations is ~ 1000 light years (note that civilisations separated by this distance require at least twice this time span i.e. 2000 years for the first initial 2-way transmission sequence to complete).

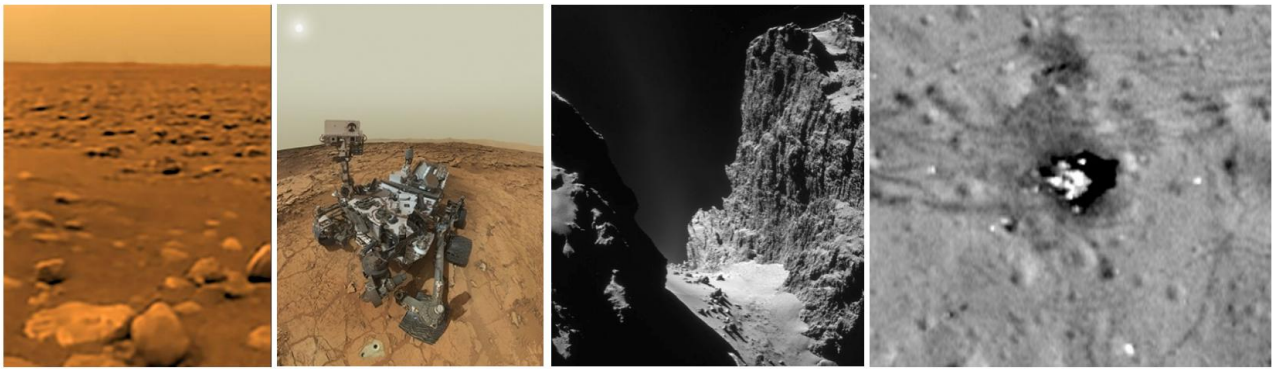

Fig.1. So far the rest of the Solar System has proved lifeless (from left to right: Titan, Mars, Comet 67/P and the Moon). Artefacts of our own space missions are easily identified (see far right: Apollo 17 landing site) but there is no evidence of visits from other advanced, space-faring civilisations.

We should reiterate that even these estimates are probably highly optimistic, given the idealistic value assumed for at least one of the unknowns ($f_{BIO} \sim 1$). For example, although the search for life in our own Solar System has only just begun, the sterile conditions so far observed on all other Solar System bodies (e.g. the Moon, Titan, Comet 67/P Churyumov-Gerasimenko, Venus and indeed Mars) does not serve to bolster the confidence of this author regarding the ubiquity of life under less than optimal circumstances. Most likely, $f_{BIO} < 1$, and perhaps much less than one.

Many objects in the Solar System (including our own Moon) have also been imaged at very high resolution – since many are geologically inert, they appear (apart from the effects of meteor impacts) largely pristine – i.e. there is no evidence to be found of visiting spacecraft (apart from our own) or any other alien artefacts. This, together with the current non-detection of the signature of other large alien artefacts in the Milky Way (e.g. Dyson spheres) [5,6,7], suggests that advanced Kardashev type II and type III civilisations [8] may not exist. The consequences of this are two-fold, first of all this may imply that the general longevity of a technically capable species is rather limited, and secondly that if there are artificial radio signals out there to be found, they may be being generated by civilisations operating with energy budgets that may not be too much more advanced than our own.

In short, we should not under-estimate the challenge of detecting artificial radio signals from other civilisations. Such signals are likely to be faint, temporally sporadic and spatially rare.

## 3. SETI and the new generation of radio telescopes

Despite the significant challenges associated with SETI research, it should be emphasised that there are also reasons to be optimistic about what the future may bring. In particular, the sensitivity of our radio telescope facilities are now about four order of magnitudes more sensitive than when the first SETI searches began in the early 1960's. In addition, significant advances in instantaneous bandwidth, and the introduction of highly capable signal processing systems have also been made.

A more recent development is a major enhancement in the instantaneous field-of-view modern radio instruments enjoy – this key parameter for SETI surveys remained relatively flat from the 1940s to 1995. By the end of the last century, the first telescopes equipped with multiple receivers located at the focal plane became operational (e.g. the Parkes multi-beam receiver system [9]), and over the last few years the re-emergence of aperture array technology operating at frequencies below 300 MHz has transformed the field-of-view of low-frequency radio telescopes (e.g. LOFAR [10]). Two-dimensional Focal Plane Arrays ("radio cameras" also based on aperture array technology) are also increasing the field-of-view of traditional dishes at cm wavelengths (e.g. the WSRT-APERTIF telescope [11]).

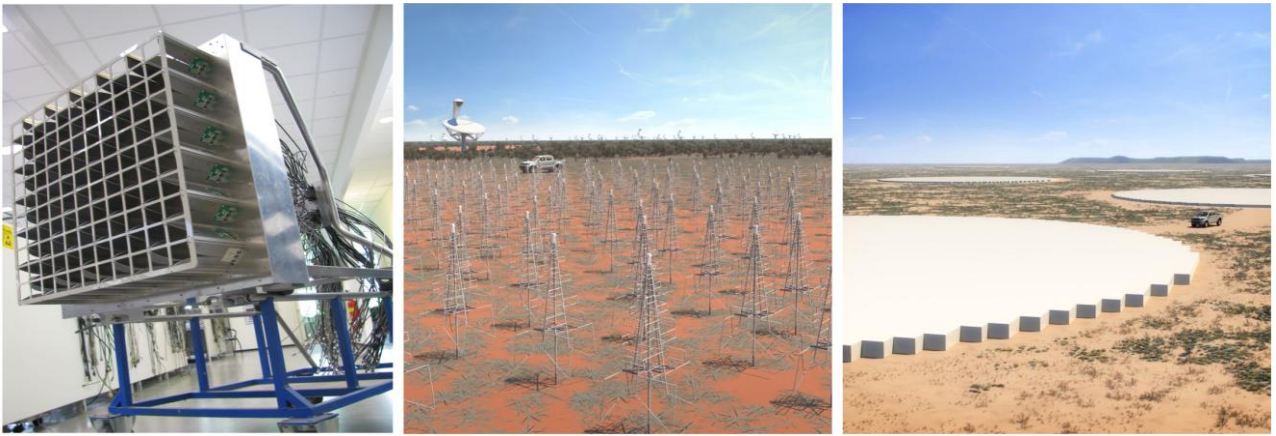

Fig.2. Aperture Array technology is a major component of SKA pathfinders such as the WSRT APERTIF telescope (left) and the SKA (artists impression centre and right).

A major step forward in terms of both sensitivity *and* field-of-view is well represented by the SKA (Square Kilometre Array). The SKA is a billion Euro global project, that will see a low frequency aperture array and higher frequency dish arrays built in the deserts of Western Australia and South Africa. The ability of arrays like LOFAR and the SKA to form multiple, independent beams in many different directions simultaneously is an important attribute to isolating SETI candidate signals from other contaminants (e.g. radio frequency interference). In principle, even phase 1 of the SKA can detect transmissions similar to an Arecibo-like planetary radar system located 300 light-years from the Earth with a coherent integration time of only 60 seconds [12]. SKA1 is expected to become operational in the first half of the next decade, this new telescope (and indeed its various pathfinders and precursors) is set poised to transform SETI research across a wide range of important parameter space but in particular, depth and field-of-view.

The SETI community is making great strides in entering into the mainstream of radio astronomy research, taking direct advantage of new and upgraded radio telescope facilities. Several significant SETI programmes have fared well in standard scientific peer-review processes, and are now being executed on some of the most sensitive radio telescopes in the world, such as Arecibo, LOFAR and the GBT [3]. The SETI case for SKA1 is already well established [12], and the goal will be to see the first commensal SETI surveys being observed early in the next decade.

## 4. SETI and Fast Radio Bursts (FRBs)

Another very important and recent development in radio astronomy with profound relevance to SETI is the progress being made in observing and understanding the transient nature of the radio sky. In particular, a new class of transient radio sources have recently emerged known as FRBs (Fast Radio Bursts) [13]. Around about a dozen of these events have now been detected, and they are characterised by their singularity, brightness (> 1

Jy), short duration (~ 1 millisecond) and very large dispersion measure (the latter being inferred via the frequency dependent ($f^{-2}$) arrival time of the broadband radio burst [14]). The discovery of this entirely new class of radio transient, clearly demonstrates the great strides being made in modern time-domain analysis of radio data.

Since the current generation of radio telescopes operating at cm wavelengths only have a very limited field-of-view, the chances of detecting a FRB are actually rather low. By extrapolation [14], it is estimated that up to 10000 FRB events occur per day across the entire sky but almost all of them currently go unobserved. Since the location of FRBs is only poorly determined by single dish positions, and since no follow-up detections have been made at other wavelengths, the nature of these events is still a matter of considerable speculation. Although no Gamma Ray Burst (GRB) events have yet been associated with FRBs, the most plausible explanations include supergiant pulses from extragalactic Neutron Stars [15] or involve the collapse/merger of degenerate compact objects also located at cosmological distances (e.g. [16]).

The consideration of whether these radio bursts might be artificial in nature seems not unreasonable – in particular, the unusually high volumetric event rate is considered large for an extragalactic population [17]. The short duration of FRB events suggests that the spatial scale associated with the phenomena is small, < 300km – clearly the kind of physical scale that any advanced technical civilisation might deal comfortably with. On the other hand, assuming the emission is largely isotropic, the energy release associated with an FRB event is considerable ~ $10^{31-33}$ Joules [14], up to several times greater than the total annual energy output of the Sun. Noting both the implied energies, event rate and cosmological distances involved, any associated civilisation is required to be of at least Kardashev Type 2. This begs the question why we don't see other evidence for the presence of such advanced civilisations elsewhere in the Universe, including in our own backyard.

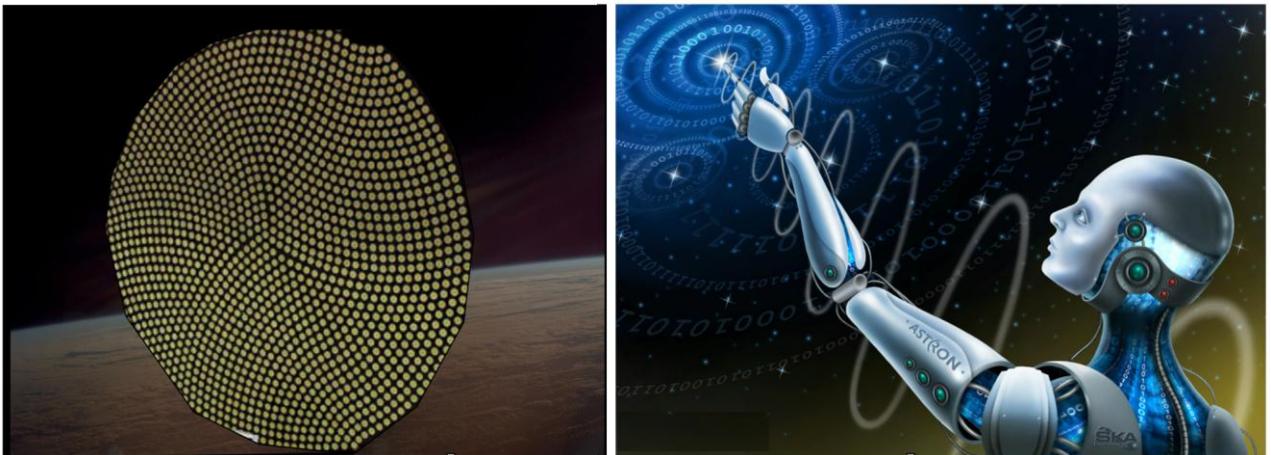

Fig.3. Left – the author's impression of a coherent ETI transmission system is an unlikely explanation for FRBs. Right – cognitive computing may represent an important leap forward for serendipitous discovery and for SETI research in particular.

The energy requirements (and thus the inferred high brightness temperature of the emission) can be relaxed by postulating some kind of beamed or coherent emission process. In the case of another technical civilisation being involved, it is not inconceivable to suggest that the radio emission is produced via a coherent system of centrally fed, transmitting antennas [18]. For example, a distributed array of transmitters with a maximum baseline of ~300 km, pointing directly towards the Earth and located ~ 1 Gpc away, requires a total power budget of ~$10^{21}$ Watts [18] (still considerable compared to our own capacity but several orders of magnitude less than that associated with a Type 2 civilisation). This energy budget can be significantly reduced by re-locating the transmitters to the confines of our own Galaxy – in this case the apparently large dispersion measures (DM) might be explained if the signal has been artificially (and negatively) chirped – a technique that is widely used in terrestrial radio communication systems, in particular with radar and spread-spectrum applications. Alternatively, perhaps the high DMs arise fairly close to the event itself e.g. via an external medium associated with the cold plasma waste products of a technically advanced civilisation. Radio emission produced by interstellar propulsion systems would not seem to fit the known characteristics of FRBs – proposed systems such as magnetic sails [19]) are not expected to be used impulsively for example, and anti-matter engines detectable in the radio would probably be even more easily detected at much higher frequencies (e.g. in the Gamma ray domain).

If FRBs are indeed some kind of directional ETI beacon or extra-galactic communication system, we would expect bursts from these locations to eventually repeat. Assuming these events occur at cosmological distances (and are therefore physically distinct and presumably operated by independent entities), some significant creativity is required to explain why the characteristics of each FRB event should be so similar to one another. The association of advanced civilisations with FRBs is currently highly speculative, and while the evidence does not in my view completely rule out such a link, it is, in all probability, unlikely. The very similar physical characteristics of FRBs detected so far, suggest that a common astrophysical explanation to this new and interesting class of cosmic transient phenomena is a more natural solution. In terms of SETI detection strategy, it should be noted that

FRBs (and probably other radio transients yet to be discovered) might represent a significant background of transient "noise" that will hardly make the identification of artificial SETI signals any easier. With 10000 FRB events predicted each day, the need to employ automated and intelligent signal extraction algorithms becomes more and more important.

**5. SETI, Big Data and Cognitive Computing**

Even the basic characteristics of an artificial signal generated by another civilisation are completely unknown to us. SETI researchers must "expect the unexpected", and presumptions about the basic nature of the signal should not be made. The raw data collected from radio telescope arrays must therefore be sampled with excellent spatial, temporal and spectral resolution, in order to preserve all possible information. This, together with a complex post-processing analysis strategy (such as coherent de-dispersion and Doppler drift searches), plainly places SETI data analysis requirements (and indeed radio astronomy in general) firmly in the realm of "Big Data" (see [20] for a more detailed description).

As large corporate investments are made in the commercial exploitation of Big Data, significant progress is being made in areas such as advanced data analytics, new visualisation methods and the introduction of advanced pattern recognition and feature extraction techniques. These can greatly benefit radio astronomy in general and perhaps SETI searches in particular. As noted earlier, the likely rate of radio transients in the era of the full SKA will reach levels that are beyond the capacity of even large, well organised teams of radio astronomers. Faced with an avalanche of data from all sides, the "brute force" approach of commercial data analytics, and especially the future promise of cognitive computing, could have a huge impact in terms of increasing the chances of serendipitous discovery – fields of research such as SETI, where human bias and other pre-conceptions may limit current efforts, stand to benefit enormously.

**6. Conclusions**

The detection of a SETI signal is challenging but certainly not impossible. The potential number of signals to be detected, their nature and strengths is largely unknown. However, substantial progress is being made in terms of the increased capabilities of new and upgraded radio telescopes, future instruments such as the SKA will continue to expand our capacity for discovery through the following decade and beyond. The recent discovery of a new type of radio transient – FRBs is another welcome development, and a sign that radio astronomers are now better equipped to detect rare SETI-like radio signals. While FRBs are unlikely to be signposts of activity associated with advanced extra-terrestrials, they represent a background of transient signals from which artificial SETI signals must be identified. It's entirely possible that additional examples of natural radio transient phenomena will be discovered over the coming years, further enriching (or complicating…) the dynamic radio sky. In order to make sense of all these events and of the current radio astronomy Big Data avalanche, taking advantage of the progress currently being made in areas such as data analytics, visualisation and automated feature extraction is extremely important. Future developments in areas such as cognitive computing may be essential in maintaining the pace of serendipitous discoveries – removing humans from part of the discovery process might actually benefit areas such as SETI in which human preconceptions may represent the ultimate "great filter".

Acknowledgements

The author would like to acknowledge the support of an IBM Faculty Award.

References

[1] G. Cocconi & P. Morrison, Nature 184 (1959) 844.
[2] F.D. Drake, 1961, Physics Today 14 (1961) 40.
[3] A.P.V. Siemion, P. Demorest, E. Korpela, et al. The Astrophysical Journal 767 (2013) 94
[4] N. Prantzos, International Journal of Astrobiology, 12, 3 (2013) 246.
[5] R. Carrigan, Acta Astronaut. 78 (2012) 121-126.
[6] J.T. Wright, R.L. Griffith, S. Sigurdsson et al. The Astrophysical Journal, 792 (2014a), 27
[7] J.T. Wright, B. Mullan, S. Sigurdsson et al. The Astrophysical Journal, 792 (2014b), 26
[8] N. Kardashev, Transmission of Information by Extraterrestrial Civilizations, Soviet Astronomy 8 (1964) 217.
[9] L. Stavely-Smith, W.E. Wilson, T.S. Bird et al. Publications Astronomical Society of Australia 13 (1996) 243-248.
[10] M.P. van Haarlem, M.W. Wise, A.W. Gunst et al. Astronomy & Astrophysics 556 (2013) A2.
[11] W.A. van Cappellen, L. Bakker and T.A. Oosterloo, Proc. XXXth URSI General Assembly (2011) 1-4.
[12] A.P.V. Siemion, J. Benford, Jin Cheng-Jin et al. Proceedings of Science, http://arxiv.org/pdf/1412.4867v1.pdf (in press).
[13] D.R. Lorimer, M. Bailes, M.A. McLaughlin et al. Science, 318, (2007) 777.
[14] D. Thornton, B. Stappers, M. Bailes et al. Science 341 (6141) (2013) 53-56.
[15] J.M. Cordes & Ira Wasserman, http://arxiv.org/abs/1501.00753 (in press).
[16] H. Falke and R. Rezzolla, Astronomy & Astrophysics 562 (2014) A137.
[17] S.R. Kulkarni, E.O. Ofek, J.D. Neill et al. Astrophysical Journal 797 (1) (2014) 70.
[18] J. Luan and P. Goldreich, Astrophysical Journal 785 (2) (2014) L26.
[19] R. Zubrin. ASP Conference Series 74 (1995) 487.
[20] M.A. Garrett, IOP Conference Series: Materials Science and Engineering, Volume 67 (1) (2014) 012017.